\documentclass{elsart}

\usepackage{epsfig}
\usepackage{amssymb}
\def\lsim{\raise0.3ex\hbox{$<$\kern-0.75em\raise-1.1ex\hbox{$\sim$}}}
\def\gsim{\raise0.3ex\hbox{$>$\kern-0.75em\raise-1.1ex\hbox{$\sim$}}}
\begin{document}

\begin{frontmatter}



\title{Properties of the Quark Gluon Plasma: \\
A lattice perspective\thanksref{label1}}


\author{Frithjof Karsch}
\thanks[label1]{This work has been authored under contract number
DE-AC02-98CH1-886 with the U.S. Department of Energy.}
\ead{karsch@bnl.gov}
\address{Physics Department, Brookhaven National Laboratory, 
Upton, NY 11973, USA}

\begin{abstract}
We discuss results from lattice calculations for a few observables that 
are sensitive to different length scales in the high temperature phase
of QCD and can give insight into its non-perturbative structure. 
We compare lattice results with perturbative calculations
at high temperature obtained for vanishing and non-vanishing
quark chemical potential. 
\end{abstract}
\end{frontmatter}
\vspace{-12.8cm}
\hfill {BNL-NT-06/35}
\vspace{12.0cm}

\section{Introduction}
\label{Introduction}

The observation of large elliptic flow and jet quenching  
at RHIC \cite{RHIC} seems to suggest a rapid thermalization of the dense 
matter created in relativistic heavy ion collisions.  
This in turn suggests that the hot and dense medium  created in these 
collisions is still strongly interacting. It is not at all an ideal gas as one
might have expected from a straightforward extrapolation of the asymptotic 
high temperature behavior of QCD all the way down to temperatures in the 
vicinity of the transition region from hadronic to partonic matter. 

Already from the early perturbative calculations it is well known 
that new non-perturbative scales arise at high
temperature, which even may invalidate a perturbative approach to some basic
physical quantities at any temperature \cite{Gale}. 
Also the early non-perturbative
studies of finite temperature QCD, eg. calculations of the equation
of state in a purely gluonic SU(3) gauge theory \cite{boyd}, have shown
that the high temperature phase of QCD is far from being simply an 
ideal quark-gluon gas; large deviations from ideal gas behavior, 
$\epsilon = 3p$, occur even at temperatures $T\sim (2-3)T_c$. 

Non-perturbative length scales characterizing the screening of electric and
magnetic modes at high temperature, which have been identified in perturbative 
calculations and have been studied quantitatively in non-perturbative
lattice calculations, have been implemented in refined perturbative 
calculation schemes. This led to the hard thermal loop resummation 
scheme \cite{pisarski} as well as the reformulation of perturbative
calculations in terms of separate
effective theories for the electric and magnetic sectors of QCD \cite{braaten}.
This allowed to incorporate systematically non-perturbative aspects of QCD 
into perturbative calculations and produced for various observables 
impressive agreement with lattice calculations at high temperature,
even at temperatures as low as $T\simeq 2T_c$.

In view of the experimental findings
at RHIC and the speculations on the nature of the interactions still
present in the QCD plasma phase at high temperature \cite{Shuryak}, 
it may be useful to
recollect some of the results obtained in lattice studies of 
observable that are sensitive to different length scales characterizing
the high temperature phase of QCD. 
We will discuss here results on screening lengths in the QGP as well as
the QCD equation of state at non-zero temperature and baryon number
density. We will compare these
results with refined perturbative calculations and, in particular, 
want to stress
that (i) most of the non-perturbative features of QCD at high temperature
and density are already present in quenched QCD, {\it i.e.} the SU(3) 
gauge theory, and that (ii) the non-zero density sector seems to show
remarkably little non-perturbative structure.

\section{Thermal and non-thermal length scales at high temperature}

Attempts to describe deviations from ideal gas behavior in terms
of high temperature perturbation theory have clearly unrevealed the 
role of different non-perturbative length scales in the QGP. 
In addition to the thermal length scale which is controlled
by the lowest non-zero Matsubara frequency, $r_{therm} \sim 1/T$, 
larger length scales related to the screening of electric modes 
(Debye screening) at distances $r_E \sim 1/gT$ and magnetic modes
at distance $r_M \sim 1/g^2T$ have been identified. Moreover, 
non-thermal hard scales that control processes at
distances smaller than $r_{\rm therm}$ or, equivalently, energies 
significantly larger than
the energy scale given by temperature, still play an important role, eg.
in the analysis of quarkonium bound states or jets
in hot and dense matter. 

\begin{figure}[t]
\epsfig{file=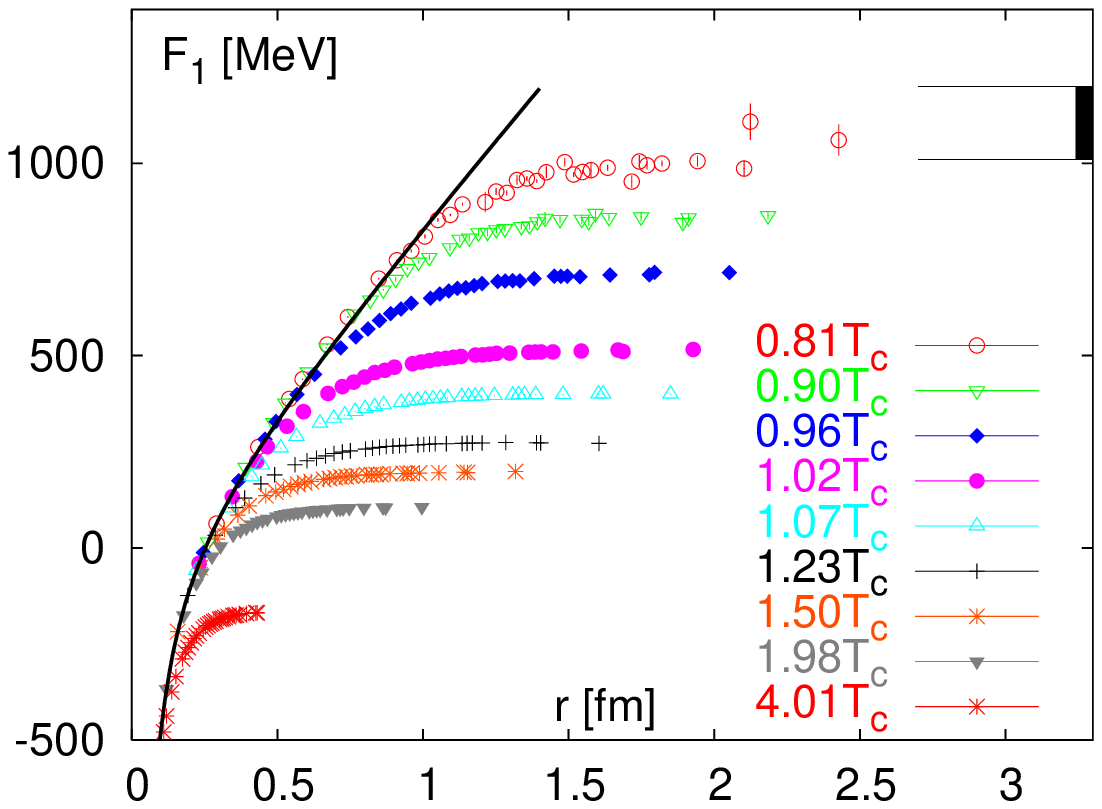,width=7.2cm}
\hspace*{-0.4cm}\epsfig{file=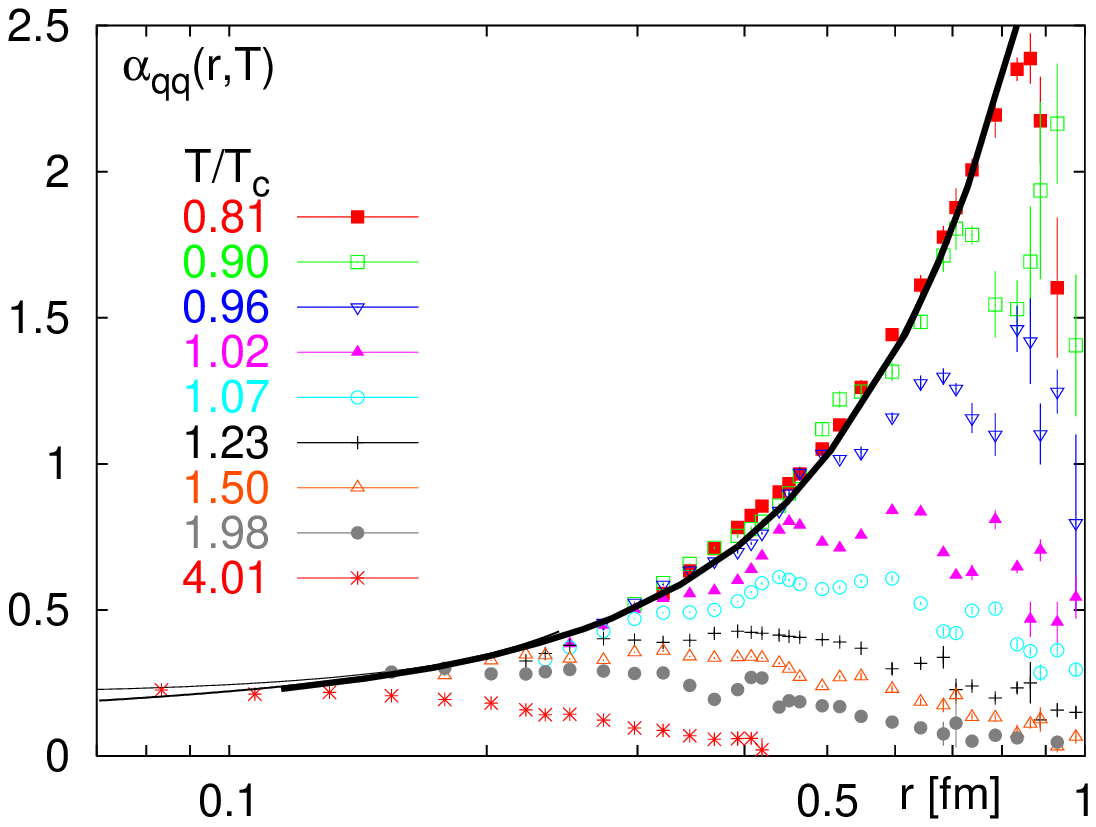,width=7.2cm}
\caption{The excess free energy of 2-flavor QCD \cite{Zantow}
induced by a pair of 
static quark antiquark sources in the color singlet channel (left) and
the running coupling (right) defined in Eq.~\ref{running}.
}
\label{fig:free}
\end{figure}

The separation of thermal and non-thermal scales becomes very
transparent from an analysis of the distance and temperature dependence 
of heavy quark free energies. While at sufficiently short distances heavy 
quark free energies are $T$ independent even at quite large $T$,
they are screened at large distances for any temperature. 
This is shown in Fig.~\ref{fig:free}(left) for a calculation in 2-flavor
QCD \cite{Zantow}. Motivated by one loop perturbation 
theory for the singlet free energy, $F_1(r,T) = -4\alpha_{q\bar{q}}/3r$ with
$\alpha_{q\bar{q}} =g^2(r)/4\pi$, one may define a running coupling,
\begin{equation}
\alpha_{q\bar{q}}(r,T) = \frac{3r^2}{4} \frac{{\rm d} F_1(r,T)}{{\rm d} r}\; .
\label{running}
\end{equation}
Fig.~\ref{fig:free}(right) shows that this coupling agrees with the 
zero temperature perturbative form of the running coupling at short 
distances and follows the zero temperature behavior even in the 
non-perturbative regime for temperatures $T\lsim 1.5 T_c$. Here
the rapid rise of the coupling with increasing distance concurs with the fact 
that the singlet free energy still follows the linear 
rising zero temperature confinement potential also in this temperature 
regime above $T_c$. This suggests that remnants of the confining force
survive the transition to the high temperature phase
and play an important role for physical processes (bound states) that are
characterized by length scale less than $r_{\rm screen}$.
We note that this
is not related to the fact that the transition in QCD with physical
quark masses is not a true phase transition but rather a rapid crossover
from the hadronic regime to the quark gluon plasma. The $T$ and 
$r$-dependence of free energies and the running coupling are similar 
also for a SU(3) gauge theory. In the high temperature phase the transition 
from non-thermal, vacuum-like
behavior to thermal screening of the free energy and the running coupling
occurs in quenched as well as 2-flavor QCD at a similar distance scale,
$r_{\rm screen}$.
To a good approximation one thus finds for the running coupling, 
\begin{equation}
g^2(r,T) \simeq g^2(r,0)\quad {\rm for} \quad 
r  < r_{\rm screen} \sim 0.5\; T_c/T \; .
\label{screen}
\end{equation}

\section{Screening of electric and magnetic modes}

{\it Debye screening:}
At high temperature the interaction among static quark sources, mediated
by gluons, is screened. The corresponding electric screening length can
be calculated in leading order perturbation theory at high temperature
and for non-vanishing quark chemical potential $\mu_q$, 
\begin{eqnarray}
\frac{m_D(T,\mu_q)}{g(T)\; T} &=& \sqrt{\frac{N_c}{3}+\frac{n_f}{6}+
\frac{n_f}{2\pi^2}\left(\frac{\mu_q}{T}\right)^2 } \;
f_E(g^2 (T),T,\mu_q) \nonumber \\
&=& m_0 (T) + m_2(T)\left( \frac{\mu_q}{T} \right)^2 +{\mathcal O} (\mu_q^4)\; .
\label{mD}
\end{eqnarray}
Here $f_E(g^2(T),T,\mu_q)$ summarizes all higher order perturbative
as well as any non-perturbative contributions. We note that $f_E$
would be a function of $g^2$ only, if only perturbative terms 
would contribute to $m_D$;
in leading order perturbation theory we have $f_E =1$.

\begin{figure}[t]
\epsfig{file=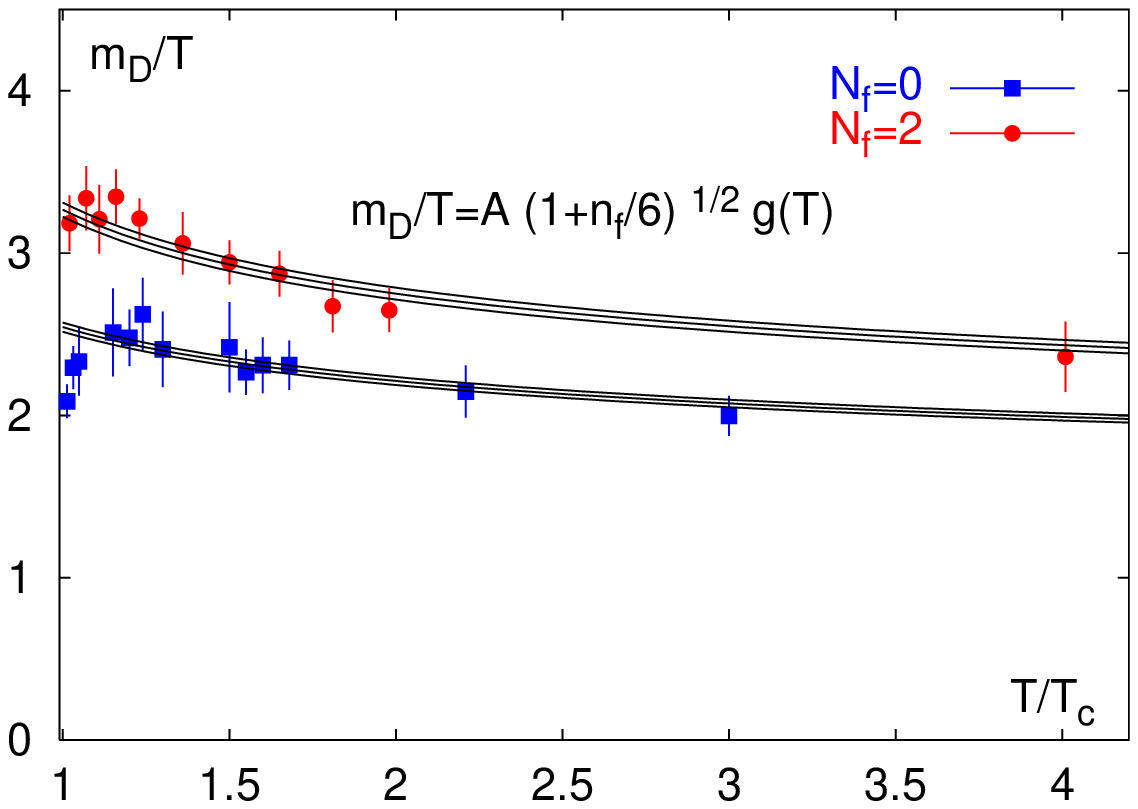,width=7.2cm}
\hspace*{-0.4cm}\epsfig{file=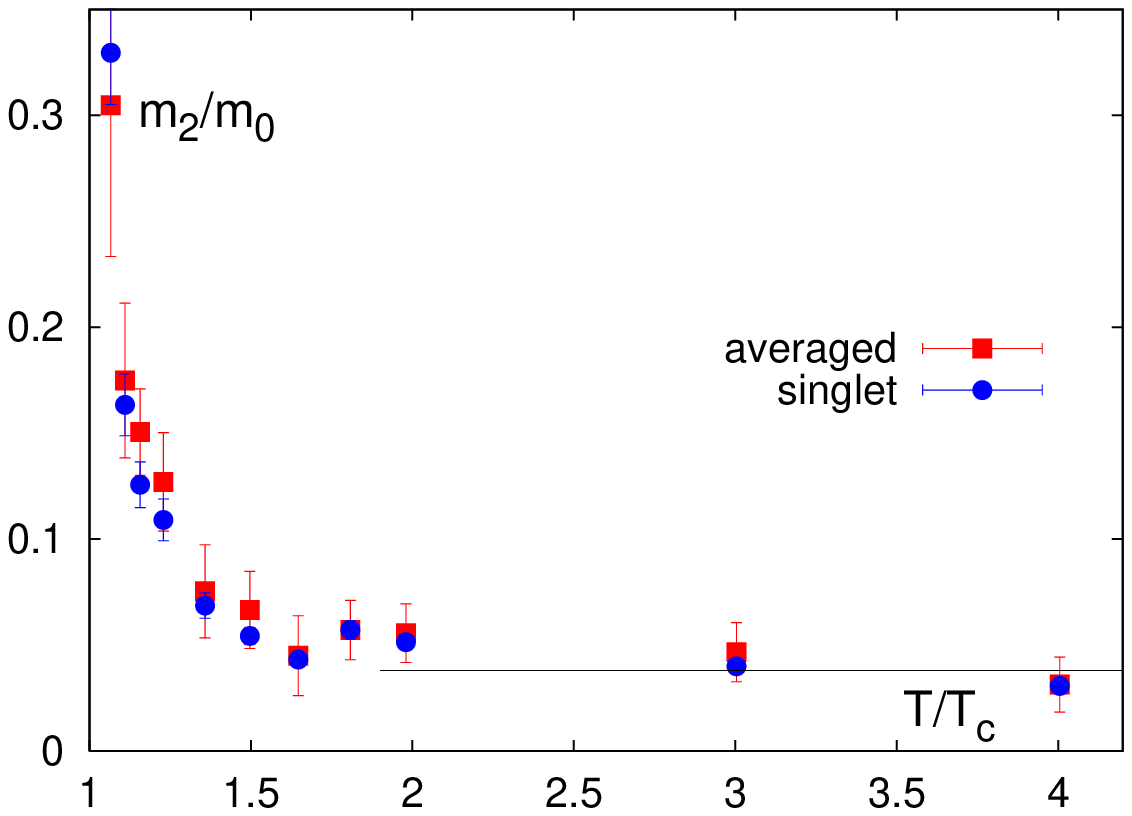,width=7.2cm}
\caption{Screening masses in 2-flavor QCD and an SU(3) gauge theory
extracted from Polyakov loop correlation functions for the singlet
free energy for vanishing chemical potential (left). The right hand 
figure shows the leading order correction for non-zero quark chemical 
potential in units of the Debye mass at $\mu_q=0$.  
The horizontal line gives the leading order perturbative result,
$m_2/m_0 = 3/8\pi^2$.
}
\label{fig:debye}
\end{figure}

In lattice calculations for the pure SU(3) gauge theory the electric 
screening mass has been calculated from Polyakov loop correlation functions 
up to fairly high temperature \cite{Kajantie}. It has been found 
there that deviations from the leading order perturbative result are 
large in the vicinity of the phase transition temperature and that the
approach to the leading order perturbative form is slow; it may 
approximately be realized only at temperatures as large as $10^{100}T_c$. 
This is in 
accordance with the logarithmic running of the QCD coupling, which in fact
is observed to be well described by the 2-loop $\beta$-function of QCD 
already close to $T_c$. In Fig.~\ref{fig:debye}(left) we show a comparison
of the electric screening mass extracted from the long distance behavior
of singlet quark-antiquark free energies in 2-flavor QCD and the $SU(3)$
gauge theory \cite{Zantow}. We note that in both cases at temperatures 
up to a few times the transition temperature the variation of 
$m_D/T$ with $T$ is well accounted for by the 2-loop running of the coupling.
Higher order corrections and non-perturbative contributions in this 
small temperature interval are well approximated by an almost flavor 
independent
constant $f_E(g^2 (T),T) \simeq A$, with $A=1.52(2)$ for quenched QCD 
and $A=1.42(2)$ for 2-flavor QCD \cite{Zantow}. 

The right hand part 
of Fig.~\ref{fig:debye} shows the leading order (quadratic) correction
to $m_D$ as function of the quark chemical potential \cite{Doring}. 
It is apparent that deviations
from the leading order perturbative result are large in the vicinity of
$T_c$. However, already at $T\simeq 1.5 T_c$ the corrections agree
well with the leading order perturbative value, which is obtained
from an expansion of the $\mu_q$-dependent prefactor in Eq.~\ref{mD}. 
The weak flavor dependence
of $f_E(g^2 (T),T)$ as well as the early onset of perturbative 
behavior for the ratio $m_2/m_0$ already at temperatures 
$T\gsim 1.5 T_c$ suggest that non-perturbative aspects of the screening
of electric modes in the high temperature phase of QCD are mainly controlled
by the gluonic sector.

{\it Spatial string tension:}
It has been pointed out already in 1980 by A. Linde that in QCD
magnetic modes are screened only at ${\mathcal O}(g^2T)$ \cite{Linde}. 
The proportionality
factor of, eg. the magnetic screening length, is entirely
non-perturbative. Nonetheless,
the modern formulation of high temperature perturbation theory takes
advantage of the existence of this new scale. It 
allows to isolated the contribution of the magnetic sector of QCD in terms
of a 3-d effective theory, which is the ordinary 3-d SU(3) 
gauge theory. A remarkable prediction of this setup is
that non-perturbative contributions to observables, which at high temperature 
are dominated by contributions of magnetic modes, are flavor independent;
a flavor dependence only enters indirectly through the flavor dependence
of the running coupling. An observable that can be used to test this
aspect of the dimensional reduction approach to QCD is the spatial
string tension.
It has, indeed, been found that spatial Wilson loops still
show area law behavior in the high temperature phase of a SU(3)
gauge theory \cite{polonyi}
and thus allow to define a spatial string tension, $\sigma_s$, that is 
sensitive to the non-perturbative magnetic screening length,
\begin{equation}
\sigma_s = \lim_{R,S\rightarrow\infty} \frac{1}{RS}\ln W(R,S) 
\sim \left[ c_M g^2(T) T f_M(g^2(T),T)\right ]^2\; .
\label{sigmas}
\end{equation}
Here $W(R,S)$ denotes a Wilson loop of size $R\times S$ in a spatial
plane of a (3+1)-dimensional lattice. The proportionality constant
$c_M$ is entirely non-perturbative. In the high temperature limit
$f_M(g^2(T),T)\rightarrow 1$ and $c_M$ should agree with the 
proportionality factor for the string tension of a 3-d SU(3) gauge theory,
$\sqrt{\sigma_3} = c_3 g_3^2$. 

It has been questioned whether
the dimensional reduction scheme, which is a central concept of 
todays perturbative and resummed perturbative calculations at high
temperature, can be applicable to QCD with light quarks \cite{Gavai}.
Light compound fermionic (mesonic) modes might introduce new light degrees
of freedom that cannot be integrated out. Moreover, large spatial Wilson
loops will not show area law behavior in the presence of light quarks
and similar to the free energy shown in Fig.~\ref{fig:free} one may expect 
that the spatial potential from which $\sigma_s$ is extracted will show 
string breaking at a certain distance scale.

\begin{figure}[t]
\hspace*{-0.4cm}\epsfig{file=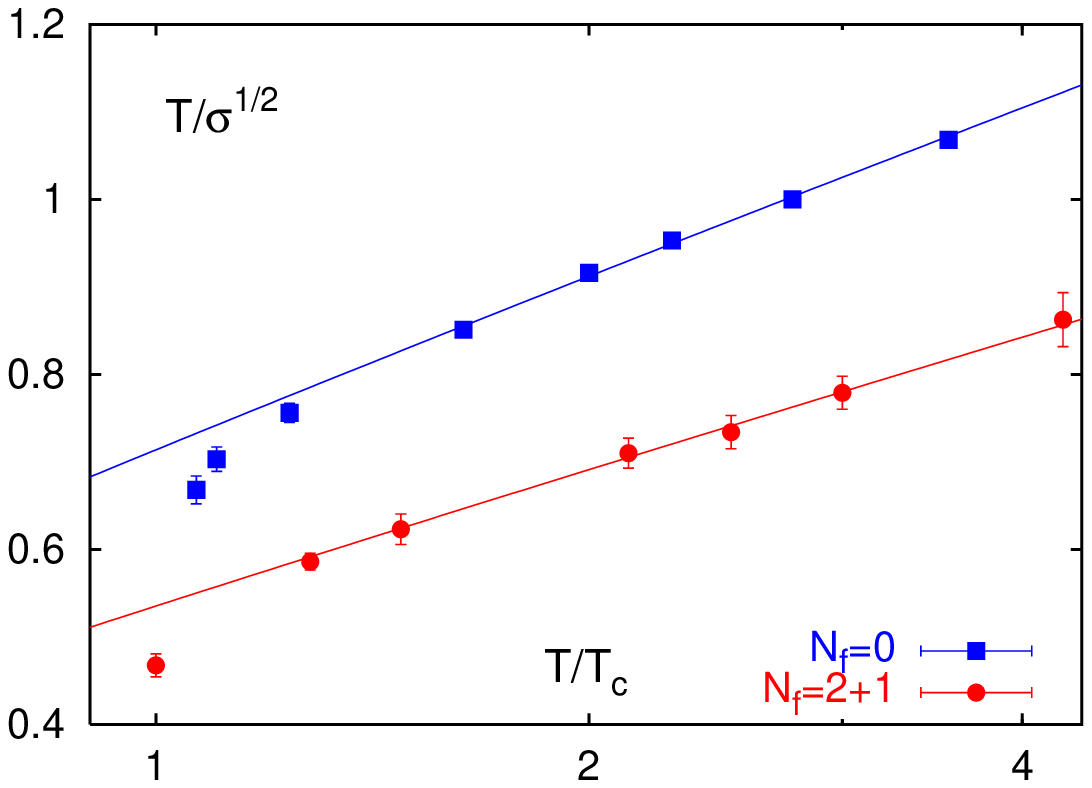,width=7.3cm}
\hspace*{-0.5cm}\epsfig{file=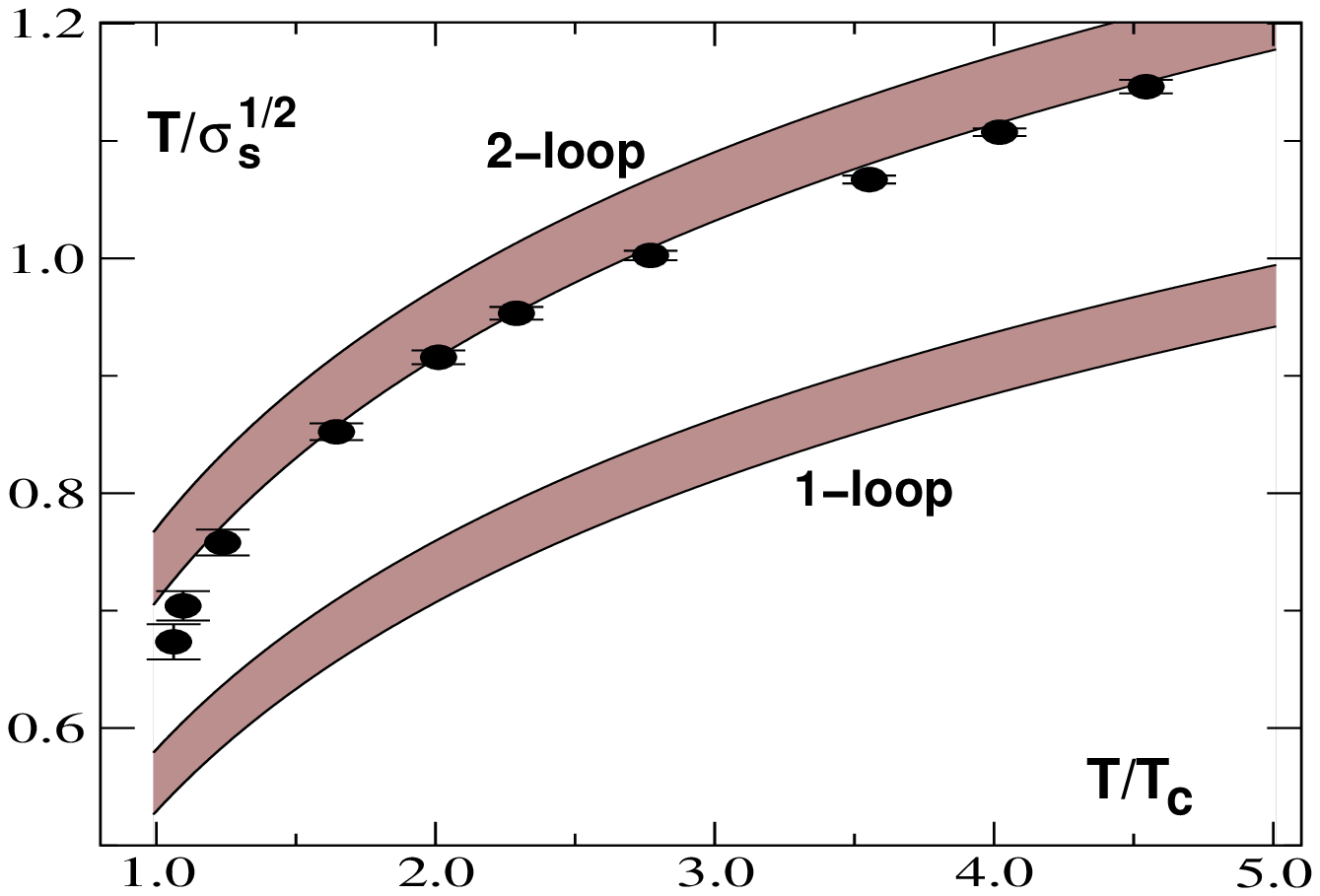,width=7.2cm}
\caption{Spatial string tension in a SU(3) gauge theory and 
(2+1)-flavor QCD (left). The right hand figure shows a comparison
of the SU(3) results with a 2-loop perturbative calculation for
the coupling in the effective 3-d theory for the magnetic sector
of QCD \cite{laine}.}
\label{fig:spatial}
\end{figure}

In Fig.~\ref{fig:spatial}(left) we show results from a calculation of  
$\sigma_s$ in quenched \cite{boyd} and (2+1)-flavor 
QCD \cite{rbcbi}. The solid lines show fits
with the leading order ansatz ($f_M\equiv 1$) given in Eq.~\ref{sigmas} 
using the perturbative 2-loop form for $g^2(T)$. The fit
yields for the proportionality constant, $c_M=0.566(13)$ for quenched QCD
and $c_M=0.587(41)$ for (2+1)-flavor QCD. This should be compared with the 
result in a 3-d SU(3) gauge theory 
\cite{Teper}, $\sigma_3 = 0.553(1)$. Fig.~\ref{fig:spatial}(right) 
shows a perturbative analysis of $\sigma_s$, which uses $\sigma_3$ as
non-perturbative leading order input \cite{laine}.
At least for this particular gluonic observable the
dimensional reduction approach thus seems to work very well down to
temperatures $T\gsim 1.7T_c$. We also note that the analysis of $\sigma_s$
yields a result for the temperature dependent running coupling $g^2(T)$.
At $T\simeq 2T_c$ one finds for quenched and (2+1)-flavor QCD, 
$g^2(2T_c) \simeq 2.0$ and $g^2(2T_c) \simeq 2.4$, respectively. This is
consistent with values deduced from the analysis of $m_D/T$ 
shown in Fig.~\ref{fig:debye}.

\section{Equation of state and perturbation theory}

\begin{figure}[t]
\epsfig{file=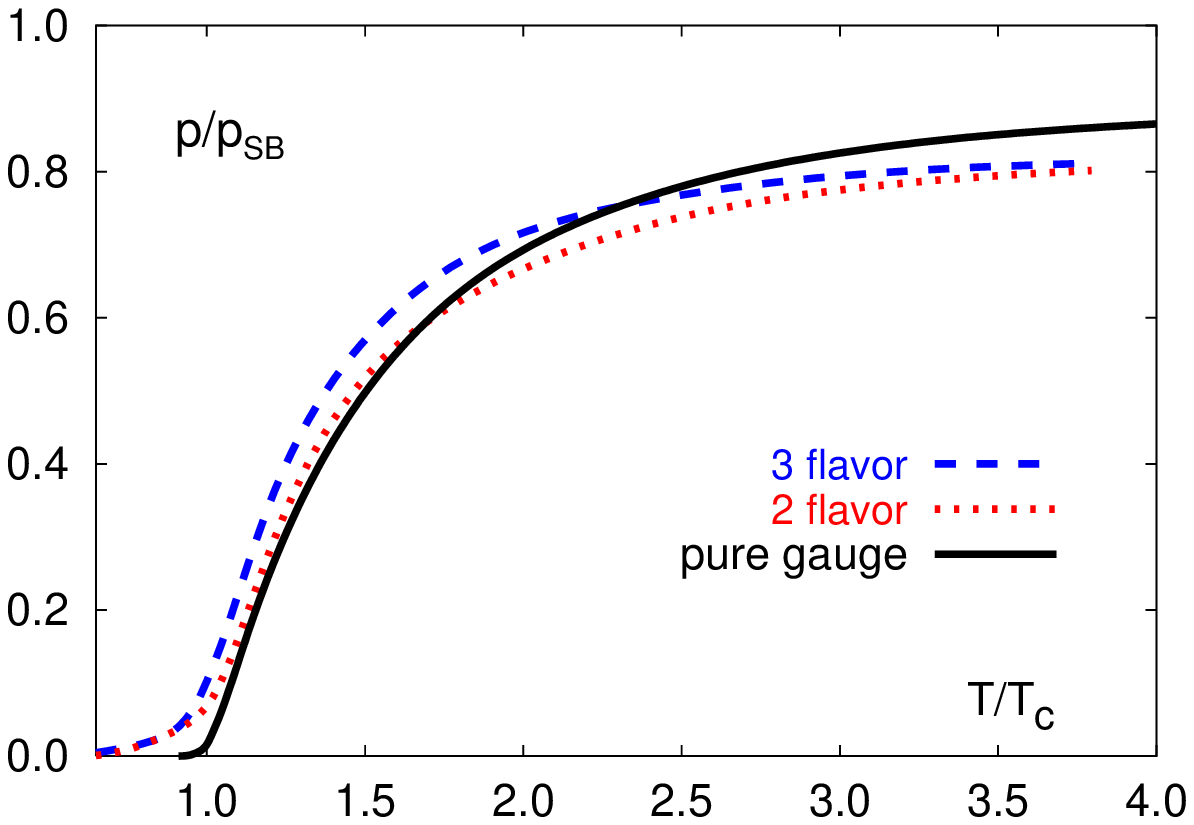,width=7.2cm}
\hspace*{-0.3cm}\epsfig{file=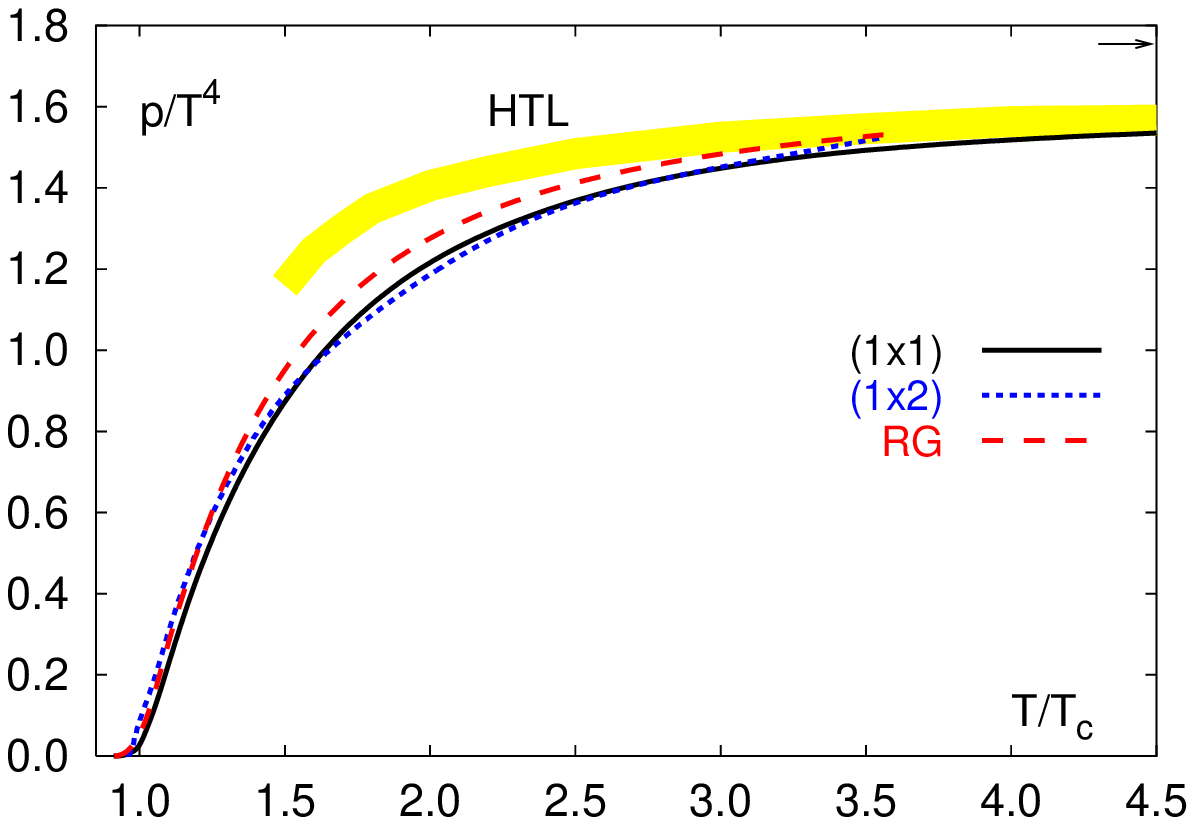,width=7.2cm}
\caption{The pressure in a SU(3) gauge theory and QCD with $2$ and $3$ 
light flavor degrees of freedom (left). The right hand part of the
figure shows a comparison of a  HTL-resummed perturbative calculation
with lattice calculations of the pressure in a SU(3) gauge theory
performed with different gauge actions.
}
\label{fig:eos}
\end{figure}

\begin{figure}[t]
\epsfig{file=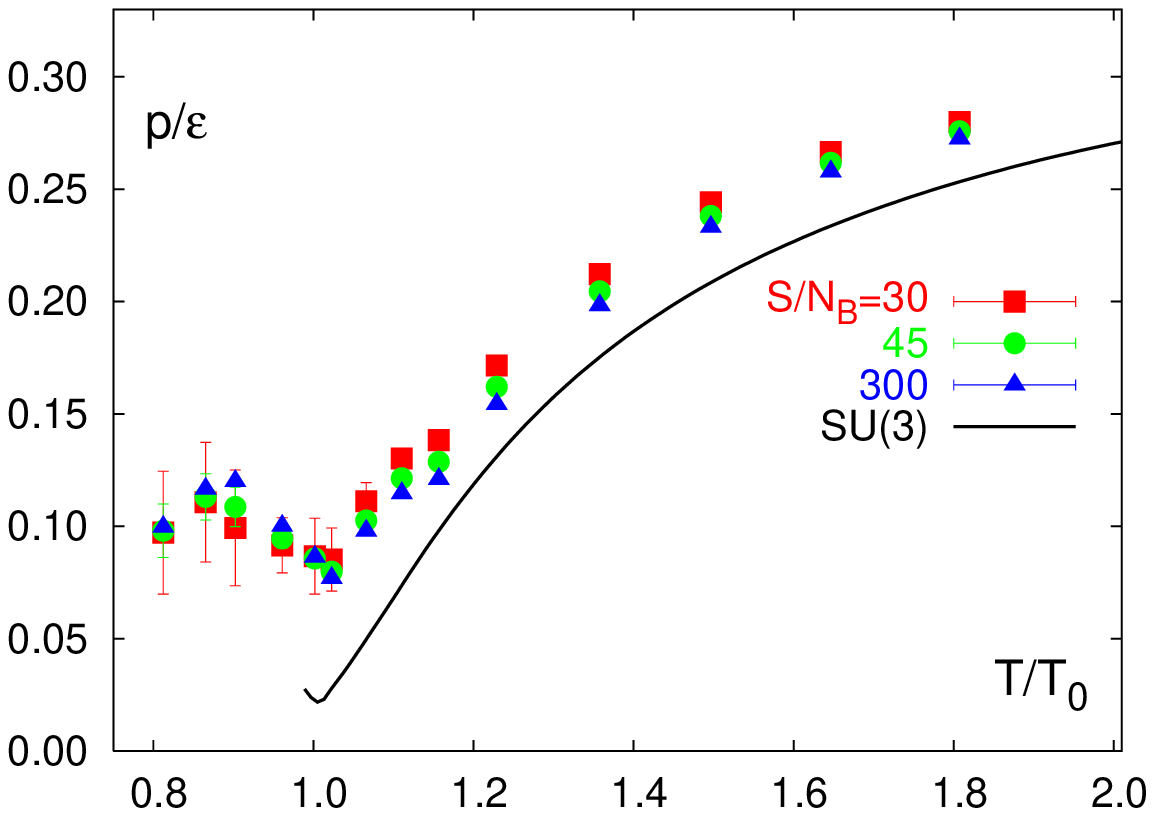,width=7.2cm}
\hspace*{-0.3cm}\epsfig{file=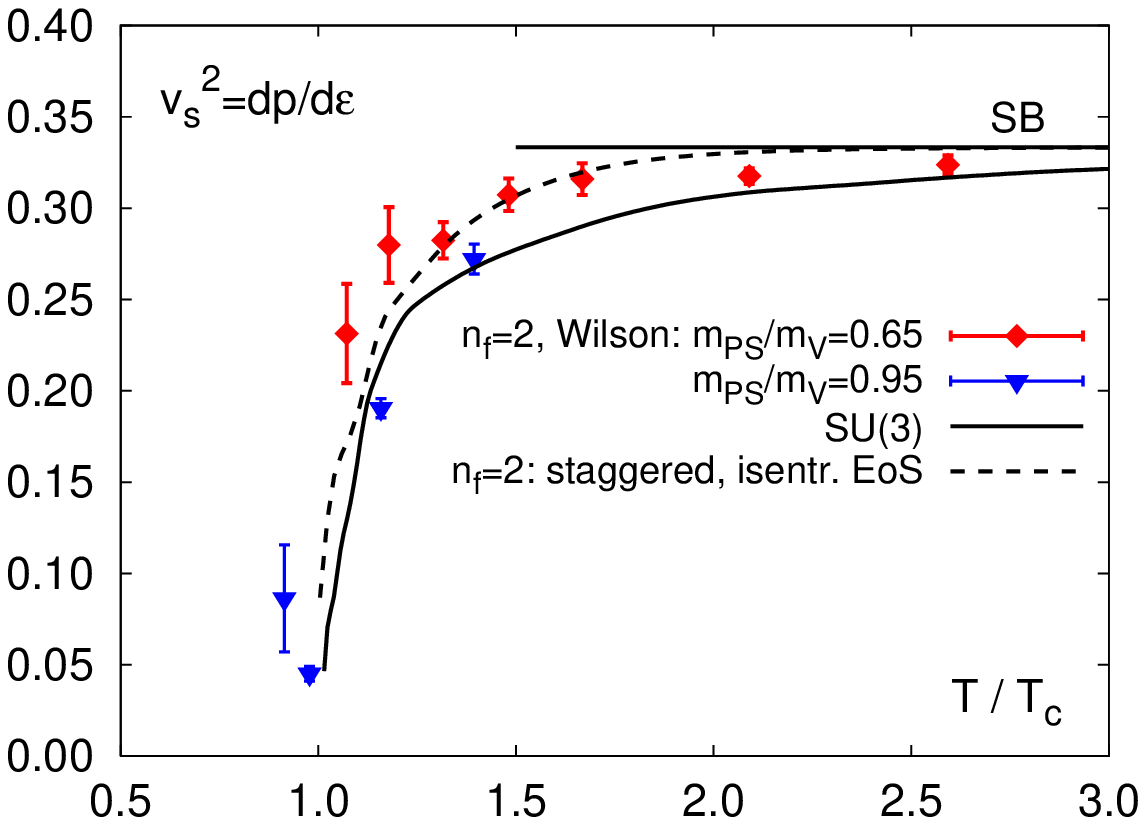,width=7.2cm}
\caption{The equation of state $p(\epsilon)$ in 2-flavor QCD
at different ratios of the entropy per net baryon number (left)
\cite{isentropic} and the velocity of sound (right).
}
\label{fig:isentropic}
\end{figure}

Probably the most direct manifestation of strong interactions in the
high temperature phase of QCD is given by the temperature dependence
of basic bulk thermodynamic observables, eg. the energy density ($\epsilon$)
and pressure ($p$). Already the early calculations in quenched QCD have 
shown that up to a few times $T_c$ $\epsilon$ and $p$ show large 
deviations from ideal gas behavior.  
E.g. one has for the pressure in QCD with 
massless quarks 
\begin{equation}
\frac{p}{T^4} =\left( 
\frac{8 \pi^2}{45} + \hspace*{-0.2cm}
\sum_{f=u,d,..} \left[\frac{7 \pi^2}{60} +
\frac{1}{2}  \left(\frac{\mu_f}{T}\right)^2
+ \frac{1}{4 \pi^2} \left(\frac{\mu_f}{T}\right)^4
\right] \right)
f_p(g^2(T),T,\mu_f) \; ,
\label{pressure} 
\end{equation} 
with $f_p\equiv 1$ and $\epsilon =3p$ in the infinite temperature limit. 
Deviations from ideal gas behavior are parametrized here by the function
$f_p$. Its structure is apparently dominated by non-perturbative
contributions arising already in the gluonic sector of QCD. This is apparent 
from Fig.~\ref{fig:eos}(left), which shows the pressure in quenched as well
as $2$ and $3$ flavor QCD with (moderately) light quarks at vanishing
quark chemical potential \cite{oldeos} and normalized to the corresponding 
ideal gas value given by the prefactor in Eq.~\ref{pressure}. The temperature
dependence of this ratio shows only little flavor dependence. Recent studies
of $\epsilon/T^4$ and $p/T^4$ with smaller quark masses and closer to the 
continuum limit \cite{MILCeos} as well as the (isentropic) 
equation of state for non-zero quark chemical potential \cite{isentropic}
(Fig.~\ref{fig:isentropic}(left))  
show that this pattern is a generic feature of QCD thermodynamics. 
In the temperature interval  $T_c\le T\lsim 3T_c$ thermodynamics thus is 
characterized by large deviations from the conformal ideal gas limit,
which also results in large deviations of the velocity
of sound, $v_s^2 = {\rm d} p/{\rm d}\epsilon$, from the asymptotic
infinite temperature value,
$v_s^2\rightarrow 1/3$ (Fig.~\ref{fig:isentropic}(right)).  

The QCD pressure for $\mu_q\ge 0$ has been analyzed to all orders that are 
calculable
in perturbation theory \cite{Vuorinen}. Non-perturbative contributions
at ${\mathcal O} (g^6)$ have been analyzed \cite{schroder} and 
the screening of electric modes has also
been implemented in self-consistent calculations \cite{blaizot}. 
We show in Fig.~\ref{fig:eos}(right) results from an analysis of the 
pressure that uses HTL-resummed gluon self energies in the gluon
propagator \cite{blaizot}. This suggests
that the structure of the pressure, in particular the deviations from
ideal gas behavior and the slow approach to the ideal gas limit 
visible for $T\gsim 3T_c$, is well understood in the perturbative 
context, by taking into account non-perturbative screening effects. 

The confrontation of refined perturbative calculations with lattice results
suggest that at least for temperatures $T\gsim (2-3)T_c$ the high temperature 
behavior of bulk thermodynamic observables indeed can be understood in 
terms of the basic partonic degrees of freedom, quarks and gluons. 
Similar conclusions can also be drawn from the analysis of hadronic
fluctuations \cite{redlich}, eg. fluctuations of the baryon 
number and correlations between baryon number and strangeness 
fluctuations.

\section{Conclusions}

We have discussed some basic results on the separation between thermal and
non-thermal length scales in the high temperature phase of QCD as well
as electric and magnetic screening lengths. 
We have shown that at short distances, $r< r_{\rm screen}\sim 0.5T_c/T$,
the QCD coupling constant is not yet modified by temperature effects.
On the other hand,
at large distance the running of the coupling is controlled by temperature;
large distance observables like the electric and magnetic screening lengths
are controlled by a temperature dependent running coupling, $g^2(T)$, which 
in magnitude is comparable
to couplings characterizing perturbative vacuum physics, {\it i.e.}
$\alpha \lsim 0.25$ or $g^2 \lsim 3$. 

Bulk thermodynamic observables, eg. energy density and pressure, show
strong deviations from ideal gas behavior even at (2-3) times the transition
temperature and approach the asymptotic ideal gas limit only slowly.
The good agreement between lattice results and refined perturbative 
calculations that take into account some non-perturbative effects,
arising from the screening of electric and magnetic modes, 
suggests that for $T\gsim (2-3)T_c$ bulk 
thermodynamics can be described in terms of quasi-particles that
carry the quantum numbers of quarks and gluons. This is further
supported by the analysis of fluctuations of conserved hadronic charges,
eg. baryon number and strangeness.

\end{document}